# Collective Evolutionary Concept Distance Based Query Expansion for Effective Web Document Retrieval

C. H. C Leung
Dept. of Computer Science
Hong Kong Baptist University
Hong Kong
clement@comp.hkbu.edu.hk

Alfredo Milani
Dept. of Mathematics and Computer Science
University of Perugia
Perugia, Italy
milani@unipg.it

Yuanxi Li
Dept. of Computer Science
Hong Kong Baptist University
Hong Kong
yxli@comp.hkbu.edu.hk

Valentina Franzoni
Dept. of Mathematics and Computer Science
University of Perugia
Perugia, Italy
valentina.franzoni@dmi.unipg.it

*Abstract*— **In this work several semantic approaches to concept-based query expansion and re-ranking schemes are studied and compared with different ontology-based expansion methods in web document search and retrieval. In particular, we focus on concept-based query expansion schemes, where, in order to effectively increase the precision of web document retrieval and to decrease the users' browsing time, the main goal is to quickly provide users with the most suitable query expansion. Two key tasks for query expansion in web document retrieval are to find the expansion candidates, as the closest concepts in web document domain, and to rank the expanded queries properly. The approach we propose aims at improving the expansion phase for better web document retrieval and precision. The basic idea is to measure the distance between candidate concepts using the PMING distance, a collaborative semantic proximity measure, i.e. a measure which can be computed by using statistical results from web search engine. Experiments show that the proposed technique can provide users with more satisfying expansion results and improve the quality of web document retrieval.**

I. INTRODUCTION

Query expansion (QE) is the process of reformulating a seed query to improve retrieval performance in information retrieval operations.[1] In the context of web search engines, query expansion involves evaluating a user's input (which words were typed into the search query area, and sometimes other types of data) and expanding the search query to match additional documents. Query expansion involves techniques such as the following:
- Finding synonyms of words, and searching for the synonyms as well
- Finding all the various morphological forms of words by stemming each word in the search query
- Fixing spelling errors and automatically searching for the corrected form or suggesting it in the results
- Re-weighting the terms in the original query

Query expansion is a widely studied methodology in the field of computer science, particularly within the realm of natural language processing and information retrieval.

Most casual users of IR systems type short queries. Recent research [3] has shown that adding new words to these queries can improve the retrieval effectiveness of such queries.

In the web document search engines, the goal of query expansion in this regard is that, by increasing recall, precision can potentially increase (rather than decrease), including in the result set pages which are more relevant (of higher quality), or at least equally relevant. With query expansion, pages having higher potential to be relevant, and that are otherwise not included, can be included.

In order to increase the precision of web document retrieval and decrease the users' browsing time, the most important task is to provide users the most suitable expanded queries quickly. Therefore, to find the closest expansion candidate concepts in web document domain, and to rank the expansion queries properly, are two main issues for query expansion in web document retrieval.

Our work mainly focuses on these two targets to improve the expansion results for better precision. In particular the use of a semantic proximity measure, the PMING distance [2, 37], is proposed and experimented.

This paper is organized as follows.
Related work on query expansion and proximity measures will be introduced in Section two; the proposed distance-based query expansion system for web document search will be presented in detail in Section three. The experimental results are reported in Section four, followed by conclusions in the last Section.

## II. RELATED WORKS

### A. Expansion techniques

In order to find the candidate concepts for query expansion in web document domain, different classes of expansion techniques can be considered.

One of the main approach to query expansion consists in using the associativity rules underlying the domain and the context of the query. For example, if a document contains two objects/concepts, say $U$ and $V$, where only $U$ is indexed, then searching for $V$ will not return the web document in the query result, even though $V$ is present in the web document but for some reasons it has not been *explicitly* indexed. On the other hand the presence of particular objects in a web document often implies the presence of other objects, then the association rule $U \rightarrow V$, can be used to infer missing index. The application of such inferences will allow the index elements $T_i$ of a web document to be automatically expanded by additional terms $T_j$. Another problem arises when more than one association rule, say $U \rightarrow V$ and $W \rightarrow V$, apply to the expansion of the query for $V$ thus producing two or more possible expansion candidates, like $U$ and $W$. In this case the expansion can be driven by some probability associated to the rules in the given context.

For example the associations
$block \rightarrow note \quad music \rightarrow note \quad footer \rightarrow note$
produce {*block, music, footer*} as candidates for the expansion of the query term *note*, in a context where the user is related to "musical instruments" the second rule appear the most appropriate, while in a context of "journalism" the other ones appear more appropriate or *probable*.

A main issue is therefore the question of how the appropriate associativity rules for a domain are discovered, evaluated and compared. A common source for discovering associative rules consists in using taxonomies or other hierarchical structures which express relationships among concepts or objects:

### 1) hierarchical expansion techniques
The hierarchical expansion techniques consist in extracting the association rules from the relationships embedded in a hierarchical knowledge structure.
Objects in a hierarchy can be classified as:
(i) Concrete, where the relevant objects are well-defined and independent (e.g. an orchestra, is an object by itself which *consists_of* conductors, violins, trumpets, clarinets et cetera which in their turn or *part_of* the orchestra, see Fig. 1)
(ii) Abstract, where the objects are not concretely defined (e.g. objects which abstract certain common characteristics, for instance *violins* and *trumpets* are abstracted by the abstract object *instruments*).

A first form of expansion technique is the *aggregation hierarchy expansion* which is based on the object/sub-object relationships, i.e. the hierarchical structure describes the aggregation hierarchy of sub-objects that constitute an object, in other words it represents the *part_of* relationship. Objects in such a hierarchy are usually *concrete objects*.

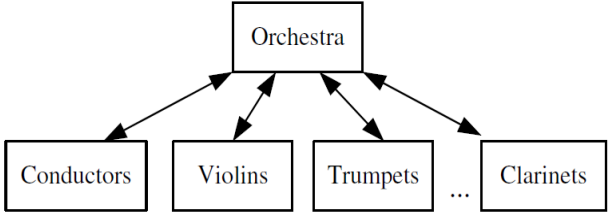

Figure 1. The concrete aggregation hierarchical expansion, where the relevant objects are well-defined. In this example, an orchestra is expanded to conductors, violins, trumpets, clarinets etc.

*Ontology-Based expansion* is a technique based on a hierarchical structure representing an ontology, i.e. objects are typically abstract and the tree represents a class/subclass hierarchy. [7]

In order to evaluate the appropriateness of the rules embedded in a hierarchical structure of index terms, a *tree traversal probability* $t_{ij}$ (Fig. 2) is associated with each branch in the hierarchy. The tree traversal probability $t_{ij}$ is defined as the probability of occurrence of the branch index $j$ given the existence of the parent index $i$. In general, the traversal probabilities of different object classes exhibit different characteristics, with $t_{ij} > t'_{mn}$ for $t_{ij}$ belonging to the concrete object class, and $t'_{mn}$ belonging to the abstract object class.

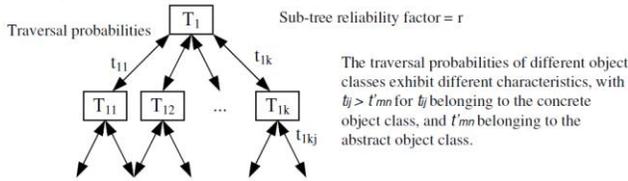

Figure 2. A tree traversal probability $t_{ij}$ which signifies the probability of occurrence of the branch index given the existence of the parent index.

*2) Co-occurrence expansion*

The *co-occurrence* based expansion tries to exploit the fact that certain semantic objects tend to occur together, despite of their hierarchical relationship. The relevant weighting is expressed as a conditional probability given the presence of other objects. An expansion to associate a web document object $O_j$ given the presence of object $O_i$ is considered significant for the expansion when

$$Prob\ [O_j\ |O_i] \geq h' \qquad (1)$$

where $h'$ is a preset threshold value that depends on the tradeoff between precision and recall performance of the system. More generally, complex probabilistic rules taking the form

$$Prob\ [O_j\ |O_1,\ldots,O_n] \geq h' \qquad (2)$$

are be applied.

In the ontology based approaches the expansion tree can be traversed bi-directionally in the course of the expansion phase. Top-down traversal will lead to an *expansion factor* > 1, while bottom-up traversal will have an *expansion factor* < 1 at each level of expansion.

There are, in general, many sub-trees whose roots are the nodes of the ontology expansion tree. Each sub-tree has been fully expanded, and it has an expansion reliability factor $0 < r < 1$, which signifies the dependability and completeness of the associated expansion. For high precision retrieval ($\pi \approx 1$), only sub-trees, which are having a significant reliability factor, need to be traversed; and nodes with a small value for $r$ will be pruned. Decision rules linking expandability with $\pi$ and $r$ can be determined. [4]

The existing ontology based query expansion methods have not been very successful especially for visual domains since ontologies are always text-oriented rather than visual-oriented and the structures of ontologies are always fixed rather than dynamic, i.e. they do not change over the time. On the other hand, for the co-occurrence expansion method, the retrieval effectiveness of the expanded queries is often no greater than, or even less than, the effectiveness of the original queries. Therefore, new query expansion methods, which provide more proper expansions, are required.

In this paper, we mainly focus on *concept distance* based query expansion for effective web document retrieval. By measuring the concept distance of queries, then ranking the expanded queries, we can provide users satisfying expansion results so that the web document retrieval efficiency is improved.

## B. Concept distance

In 2012 a new collaborative proximity measure was presented, named *PMING distance*, [2] which uses the statistical information returned by search engines as a natural source of semantics. The general idea is to use the number of occurrence of a term or a set of terms. An approximation of the number of occurrences of a term is usually provided by most search engines which return, for example, the number of results/documents which contain the query search terms.
A distance based on such number of occurrences is collaborative and dynamic, since it is based on automatic indexing of documents provided in the web by the users, and it can change dynamically over the time as new documents are indexed and old documents disappear from the web.
The PMING Distance is based on Pointwise Mutual Information (PMI) and Normalized Google Distance (NGD), which have been found to be among the measures which have a good performance in capturing the semantic information for clustering, ranking and extracting meaningful relations among concepts.
In order to understand the PMING we briefly introduce PMI, NGD and some notational conventions.

Let define:
- $f(x)$ as the number of occurrence of a search term *x* in the query results, where *x* is either a single term *t1* or a group of terms;
- $f(x,y)$ as the occurrence of two search term *x* and *y* in the query results, if *x=t1* and *y=t2* then $f(x,y)$ will count the occurrences of *t1* AND *t2* ;
- $P(x) = \frac{f(x)}{N}$. It summarizes the frequency based approach to probability, i.e. states that in the following formula probability *P* can be computed from frequency *f* and viceversa whenever the total *N* is known or can be approximated.

### 1) Pointwise Mutual Information (PMI)

*Mutual Information (MI)* is a measure of the information overlap between two random variables.
*Pointwise Mutual Information (PMI)* [25, 27] is a point-to-point measure of association, which represents how much the actual probability of a particular co-occurrence of events differs from what we would expect it to be on the basis of the probabilities of the individual events and the assumption of independence. Even though PMI may be negative or positive, its expected outcome over all joint events (i.e., PMI) is positive.
PMI is used both in statistics and in information theory.[9]
PMI between two particular events $w_1$ and $w_2$, in this case the occurrence of particular words in Web-based text pages, is defined as follows:

$$PMI(w_1, w_2) = log_2 \frac{P(w_1, w_2)}{P(w_1)P(w_2)} \quad (3)$$

This quantity is zero if *w1* and *w2* are independent, positive if they are positively correlated, and negative if they are negatively correlated.
On particularly low frequency data, PMI does not provide reliable results. Since PMI is a ratio of the probability of $w_1$, $w_2$ together and $w_1$, $w_2$ separately, in the case of perfect dependence, PMI will be *0*.
PMI is a bad measure of dependence, since the dependency score is related to the frequency of individual words.
PMI could not always be suitable when the aim is to compare information on different pairs of words.

### 2) Normalized Google Distance (NGD)

*Normalized Google Distance (NGD)* [22] is proposed to quantify the extent of the relationship between two concepts by their correlation in the search result from a search engine (e. g. Google) when querying both concepts and is defined as follows:

$$NGD(x,y) = \frac{max\{\log f(x), \log f(y)\} - \log f(x,y)}{\log M - min\{\log f(x), \log f(y)\}} \quad (4)$$

The concept of the NGD is derived from the Kolmogorov complexity, information distance and Kraft inequality, with some algebraic manipulation which involve compression-based steps that lead to normalized distances. E. g. Normalized

Compression Distance (NCD), which is a family of compression functions parametrized by a given data compressor, and its limiting case Normalized Information Distance (NID), where the number of bits in the shortest code that can be decompressed by a general purpose computable decompressor [22].

The NGD assumes that a priori all the web pages are equiprobable and tries to indirectly use the probabilistic information of the events of occurrence of a word to determine a prefix code, because the events can overlap and hence the summed probability of the *Google code* exceeds *1*. The Google code, which represents the shortest expected prefix-code word length of an event, is than approximated in the NGD.

The parameter *M*, which represents the total number of pages indexed by the search engine, which often is not known, or is varying in a short time lapse, can be approximated using any value reasonably greater than any *f(x)*.

Even if NGD is a good proximity measure, and disregarding its name, it is neither a metric, nor a distance. In fact, the NGD does not respect the property of triangular inequality.

3) *PMING Distance*

*PMING Distance* [2] consists of NGD and PMI locally normalized, with a correction factor of weight $\rho$, which depends on the differential of NGD and PMI. In PMING the two component measures are locally normalized, so that their weighted combination is based on the context of evaluation, such as on the *Vector Space Model (VSM)*.

More formally, the PMING distance of two terms *x* e *y* in a context *W* is defined, for *f(x)≥f(y)*, as a function *PMING:W×W→[0,1]*:

$$PMING(x,y) = = \rho \left(1 - \log \frac{f(x,y)M}{f(x)f(y)\mu_1}\right) + (1-\rho)\left(\frac{\log f(x) - \log f(x,y)}{(\log M - \log f(y))\mu_2}\right) \quad (5)$$

where:
- $\rho$ is a parameter to balance the weight of components ($\rho$=0.3 in our tests);
- $\mu_1$ e $\mu_2$ are constant values which depend on the context of evaluation, and are defined as:
  - $\mu_1 = \max PMI(x,y)$, with $x, y \in W$
  - $\mu_2 = \max NGD(x,y)$, with $x, y \in W$

PMING was found to incorporates the advantages of both PMI and NGD, outperforming state-of-the-art proximity measures in modeling contexts, modeling human perception and clustering of semantic associations, regardless of the search engine/repository, see Table III, Table IV and Table V for a comparison between the ranking induced by PMING and different proximity measures in contexts.

III. CONCEPT DISTANCE-BASED QUERY EXPANSION WEB DOCUMENT SEARCHING SYSTEM

A. *System Architecture*

Our web document search query expansion system mainly contains two parts: User Interface (UI) and Collaborative Query Expansion System (CQES). The system architecture is illustrated in Fig. 3. In User Interface part, we can let users type their search terms to the system like the most common web document search engine interfaces. The UI also provides the expanded queries to the user after the processing of the Collaborative Query Expansion System so that the user can choose the expanded queries from the recommended options. The User Interface will then pass to the web document search engine the expanded queries chosen by the user.

When the Collaborative Query Expansion System receive the user's original query, the system matches the submitted query with the ones in our database. If the existing query is in our database, we use it directly to find the candidate expansions; if it does not exist in our query database, we will record this new query in the system while searching for the most similar query to replace the original one. Then the expansion candidates are decided using the Query Pool. When the expansion candidates have been chosen, the distance between the matched query and the candidate queries is calculated by the Concept Distance, i.e. PMING Distance, which provide the proximity measure between the matched query and each expanded query. Next step is to rank the expanded queries according to the calculation results and turn the results back to the User Interface. The ranked expanded queries are then returned to the user.

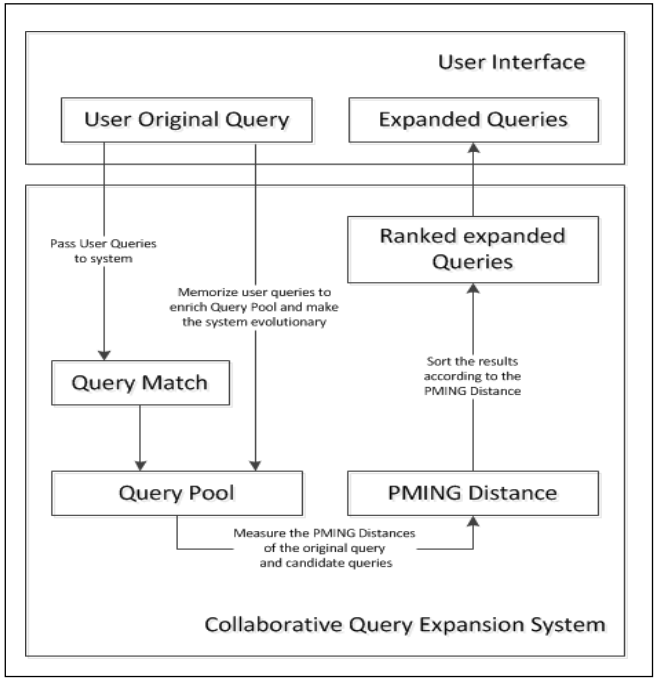

Figure 3. Query Expansion System Framework

*B. Query Pool*

The use of WordNet for query expansion has been studied by the information retrieval community [5, 6]. Mostly, this comprised retrieval of textual documents. In order to use semantic relations for retrieval of textual documents, the correct WordNet sense has to be assigned to words in the text. This is called word sense disambiguation (WSD).

WordNet [11] is one of these applications of semantic lexicon for the English language and is a general knowledge base and commonsense reasoning engine.

The purpose of the work is both to produce a combination dictionary-and-thesaurus that is more intuitively usable, and to support automatic text analysis and artificial intelligence applications.

For example, by using WordNet, 'downtown' has been expanded to 'business district', 'commercial district', 'city center' and 'city district', while 'city district' has been expanded to 'road', 'building', 'architecture', 'highway' and 'hotel'. The semantic knowledge is hierarchically expansible from the query terms and concepts and knowledge can be expanded extensively. The more extensive and complete such hierarchies, the greater the scope for rich semantic manipulation.

Recent research [8] on the topic in computational linguistics has emphasized the perspective of semantic relatedness of two lexemes in a lexical resource, or its inverse, semantic distance.

The first line of research [], which brings together ontology and corpus, tries to define the similarity between two concepts $c_1$ and $c_2$ lexicalized in WordNet, named WordNet Distance (WD). It indicates by the information content of the concepts that subsume them in the taxonomy. Formally, define:

$$sim(c_1, c_2) = \max_{c \in S(c_1, c_2)} [-\log p(c)] \qquad (6)$$

where $p(c) = \dfrac{\sum_{w \in W(c)} count(w)}{N}$

and N is the total number of nouns observed. And $S(c_1, c_2)$ is the set of concepts that subsume both $c_1$ and $c_2$. Moreover, if the taxonomy has a unique top node, then its probability is 1. In practice, we often measure word similarity rather than concept similarity. Using $s(w)$ to represent the set of concepts in the taxonomy that are senses of word $w$, define

$$sim(w_1, w_2) = \max_{c_1, c_2} [sim(c_1, c_2)] \qquad (7)$$

where $c_1$ ranges over $s(w_1)$ and $c_2$ ranges over $s(w_2)$.

It defines two words as similar if near to one another in the thesaurus hierarchy. For example, refer to Fig. 4, 'entity' can expand to 'inanimate-object'. Then 'inanimate-object' can expand to both 'natural-object' followed by 'artifact'. It then expands to both 'enclosure' and 'surface'. The former can expand to 'cage' while the latter expands to 'skin'.

WordNet [11] is a large lexical database of English. Nouns, verbs, adjectives and adverbs are grouped into sets of cognitive synonyms (synsets), each expressing a distinct concept.

Synsets are interlinked by means of conceptual-semantic and lexical relations. The resulting network of meaningfully related words and concepts can be navigated with the browser.

The most frequently encoded relation among synsets is the super-subordinate relation (also called hyperonymy, hyponymy or ISA relation). It links more general synsets like {furniture, piece_of_furniture} to increasingly specific ones like {bed} and {bunkbed}. Thus, WordNet states that the category furniture includes bed, which in turn includes bunkbed; conversely, concepts like bed and bunkbed make up the category furniture. All noun hierarchies ultimately go up the root node {entity}.

Hyponymy relation is transitive: if an armchair is a kind of chair, and if a chair is a kind of furniture, then an armchair is a kind of furniture. WordNet distinguishes among Types (common nouns) and Instances (specific persons, countries and geographic entities). Thus, armchair is a type of chair, Barack Obama is an instance of a president. Instances are always leaf (terminal) nodes in their hierarchies.

Meronymy, the part-whole relation holds between synsets like {chair} and {back, backrest}, {seat} and {leg}. Parts are inherited from their superordinates: if a chair has legs, then an armchair has legs as well. Parts are not inherited "upward" as they may be characteristic only of specific kinds of things rather than the class as a whole: chairs and kinds of chairs have legs, but not all kinds of furniture have legs.

Verb synsets are arranged into hierarchies as well; verbs towards the bottom of the trees (troponyms) express increasingly specific manners characterizing an event, as in {communicate}-{talk}-{whisper}. The specific manner expressed depends on the semantic field; volume (as in the example above) is just one dimension along which verbs can be elaborated. Verbs describing events that necessarily and unidirectionally entail one another are linked: {buy}-{pay}, {succeed}-{try}, {show}-{see}, etc. Other dimensions are speed (move-jog-run) or intensity of emotion (like-love-idolize). Adjectives are organized in terms of antonymy. Pairs of "direct" antonyms like wet-dry and young-old reflect the strong semantic contract of their members. Each of these polar adjectives in turn is linked to a number of "semantically similar" ones: dry is linked to parched, arid, dessicated and bone-dry and wet to soggy, waterlogged, etc. Semantically similar adjectives are "indirect antonyms" of the contral member of the opposite pole. Relational adjectives ("pertainyms") point to the nouns they are derived from (criminal-crime).

There are only few adverbs synsets in WordNet (hardly, mostly, really, etc.) as the majority of English adverbs are straightforwardly derived from adjectives via morphological affixation (surprisingly, strangely, etc.). The subnet derived from the adverbs is hence smaller than the other three, formed by the synsets of nouns, verbs and adjectives.

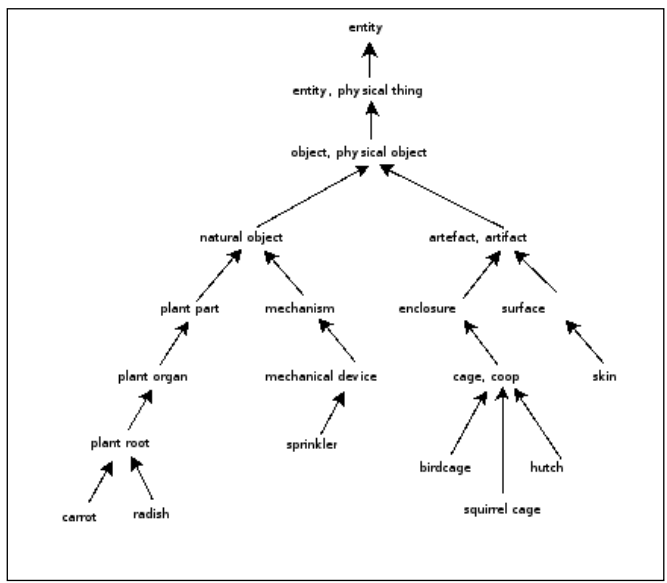

Figure 4.  WordNet "is a" relation example

We use the WordNet to provide us the expansion candidates. Then we calculate the distance between each of the candidates and matched query using Concept Distance.

## IV. EXPERIMENTAL RESULTS

In the experimental phase, both WordNet and Bing search engine were used to extract a pool of candidate terms for the query expansion. The candidate terms were then submitted as input terms to the Bing search engine, and proximity measures were calculated on the results. The PMING Concept Distance was used to calculate the affinity between the context term and each expansion candidate. Then we ranked the results according to the distances of the calculation and compared them to the users' expectation.

Fig. 5 shows an example of expansion Candidate Pool from WordNet.

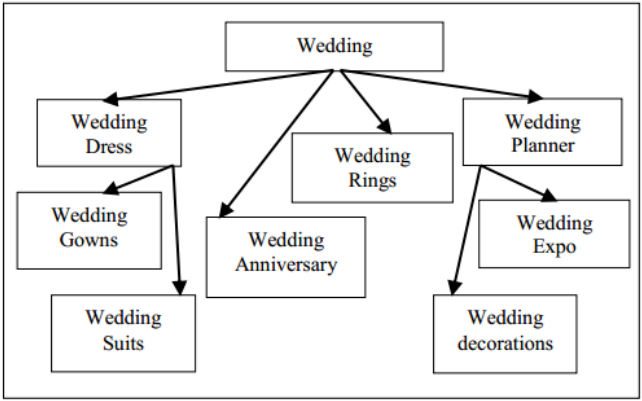

Figure 5. Query Expansion Candidate Pool Example

Table 1 shows an example of the comparison between our web document search query expansion system and Bing Web document search engine query expansion, as well as user ranked similarity of the concepts.

TABLE I AN EXAMPLE OF THE COMPARISON

| Wedding | | | | |
|---|---|---|---|---|
| WordNet Expansion | PMING Distance | Bing Search | PMING Distance | User Expectation Rank (UER) |
| dress | 0,4795552 | planner | 0,4795552 | planner |
| gown | 0,5021484 | gown | 0,5021484 | dress |
| suits | 0,4978767 | dress | 0,4978767 | rings |
| anniversary | 0,5575022 | decoration | 0,5575022 | gown |
| rings | 0,4705942 | dj | 0,4705942 | suits |
| planner | 0,4294843 | expo 2013 | 0,4294843 | decoration |
| decoration | 0,5219288 | expo | 0,5219288 | anniversary |
| expo | 0,7919295 | | | expo |
| | | | | expo 2013 |
| | | | | dj |

As we see in Table 1, the PMING Distance columns show the calculated concept distance between the pairs of Wedding and each of the expansion candidates, which were extracted both from WordNet, with a depth-first approach, and from Bing Web document search engine as further feed of candidate terms.

We have involved a group of users to test and provide the subjective results, with respect to an ideal expansion. All the expansion terms in our WordNet-Based expansions and Bing Search-Based expansions are provided to users, which are asked to rank them from the most related concepts to the least related ones according to the user personal opinion. The average ranked results are shown in the column User expectation Rank.

The users ideal expansion in User Expectation Ranks is used as a ground truth to compare the system performances.

We observe that the sorting of the candidate terms for the expansion provided by PMING Distance are ranked higher than the ones provided by the original WordNet-Based and Bing Search-Based sorting.

The same experiment has been conducted on several datasets, the observed results confirm that Concept Distance-Based query expansion greatly improves the Bing Based Query expansion, i.e. the sorting induced by Bing Web document Search

Engine. The Web document retrieval efficiency has been improved and the retrieval precision on first page turned to user has been increased.

As a critical step for the evaluation, a special care was put in the construction of the User Expectation Rank.

For each experiment, individuals among students of the Hong Kong Baptist University, University of Perugia and third party Facebook users from Italy were asked to rank the candidate terms in the English language, to build an ideal query expansion. After the voting session, the eventual typing errors and noise were removed from the results, [21] to have twenty valid rankings. The results were then passed to analysis phase.

A further result is that, comparing to PMI and NGD, the PMING Distance is confirmed to have generally better performance. As an example, we report in Table 2 the comparison of the ranking obtained using the three measures on the expansion example of Table 1, with the related values and ranking.

## V. CONCLUSION

### A. Main results

This work presents an innovative model for collaborative query expansion which is based on two main elements: a collaborative and evolutionary indexing method, and query expansion ranking method based on the PMING distance.

The architecture of an innovative system which acts as collaborative interface between the user and a search engine is also presented. The aim of the system is to improve the quality of user search experience by automatically proposing to the user a set of candidates for query expansion which are sorted by decreasing relevance, from the most relevant to the lowest one. The quality of the system is measured by two main parameters for the degree of user acceptance, i.e. the distance between the user's actual choice and the system query expansion proposal, and the quality of document retrieval, which is measured by the evaluation of classical indicators such as precision of the retrieved document and recall, combined with a qualitative measure of user satisfaction, which can be automatically detected by the number of query results which are converted into clicks .

TABLE II COMPARISON OF PMI, NGD, PMING DISTANCE

| *Wedding* | | | | | |
|---|---|---|---|---|---|
| **UER** | **PMING Ranking** | **PMI** | **PMI Ranking** | **NGD** | **NGD Ranking** |
| planner | planner | 1,8080 | gown | 0,3757 | planner |
| dress | rings | 1,6841 | planner | 0,4017 | dress |
| rings | dress | 1,5094 | suits | 0,3877 | rings |
| gown | suits | 1,7418 | decoration | 0,4377 | suits |
| suits | gown | 1,8701 | rings | 0,4593 | anniversary |
| decoration | decoration | 1,6867 | dress | 0,4553 | decoration |
| anniversary | anniversary | 1,3774 | anniversary | 0,4509 | gown |
| expo | expo 2013 | 1,0890 | expo 2013 | 0,6318 | expo 2013 |
| expo 2013 | expo | 1,0019 | expo | 0,6437 | expo |
| dj | dj | 0,6127 | dj | 0,7255 | dj |

Experiments conducted on the proposed system architecture have been presented and discussed.

The collaborative and evolutionary indexing allows new terms to increase the index as the user choose them more frequently. While the query ranking method uses the innovative PMING Distance which combines the most performing features of Pointwise Mutual Information measure and Normalized Google Distance in a single weighted proximity measure for measuring the distance between words, experiment shows that the PMING Distance better reflects the human judgment about words proximity, than its component measure taken alone. It is then used to rerank the query expansion candidates which are proposed to the user.

As shown from the results of the experiments, the accuracy of web document searching has been improved with the proposed approach of the Concept Distance-Based query expansion semantic web document search and re-ranking model. By the systematic analysis of user submitted queries, the query have been refined and enriched with knowledge base and dynamic concept distance, as well as the user participating interactive feedback, web documents could be discovered with much shorter time and higher precision. The semantic meanings and concepts of web documents are significantly enriched.

## B. Future Research Direction

Additional refinements to the current system are possible without any doubt and they are desirable in the future in order to further increase the user satisfaction. The actual results indicate the feasibility of the proposed model and system architecture.

Future research activity will focus on improving the current system implementation together with more systematic experimentation and extension of the query expansion model.

System optimization will mainly consists of minimizing the processing time among the user interface and concept distance calculation, in order to produce a faster query expansion ranking proposal to the user.

The proposed model still present several points of potential improvements. Although the human ranking scoring is performing quite well in the presented experiments, a wider range of experiments is needed both for tuning the PMING parameters with respect to the search engine which is used to determine the frequency and probability estimation of the terms whose proximity is measured. The generation of query expansion candidates, i.e. the expansion candidates pool, can also be greatly improved with different techniques, such as taking into account the user query history or other user ongoing activities taking place on the user terminal. A further innovative and very promising approach for the generation of query expansion candidates would be to use geo-located information and the history of user mobility as a source of associative rules, in this case techniques of plan inference could be used to understand the user underlying goals, such as travel destinations, expected activity (e.g. shopping, dining, working etc.) etc. which depend on the user current and previous location history. He vector space model can be used in semantic VSM models of emotions.[10][11].

This models may also be used for ranking link prediction candidates in social or other networks,

TABLE III COMPARISON OF DIFFERENT RANKING IN THE CONTEXT MOON BETWEEN CHI-SQUARE, PMI AND NGD WITH VECTOR SPACE MODEL (VSM) FROM [6]

| $Chi^2(VSM)$ | PMI(VSM) | NGD(VSM) |
|---|---|---|
| **moon** | **moon** | **moon** |
| wolf | **saturn** | wolf |
| **space** | wolf | **space** |
| **venus** | **venus** | **saturn** |
| **saturn** | horses | horses |
| horses | **space** | snake |
| bears | donkey | **venus** |
| snake | snake | donkey |
| turtle | dolphin | dolphin |
| whale | bears | bears |
| dolphin | turtle | turtle |
| donkey | whale | spiders |
| sharks | spiders | whale |
| spiders | sharks | sharks |
| bowling | bowling | bowling |
| football | golf | golf |
| volleyball | tennis | football |
| golf | football | volleyball |
| softball | baseball | tennis |
| tennis | soccer | softball |
| baseball | basketball | baseball |
| basketball | volleyball | soccer |
| soccer | softball | basketball |

TABLE IV Comparison of ranking in the context Moon between simple PMING Distance and PMING Distance with Vector Space Model (VSM) from [6]

| PMING | PMING(VSM) |
|---|---|
| **moon** | **moon** |
| *wolf* | *wolf* |
| **venus** | **space** |
| **saturn** | **saturn** |
| *horses* | *horses* |
| **space** | **venus** |
| *bears* | *snake* |
| *whale* | *donkey* |
| *snake* | *dolphin* |
| *turtle* | *bears* |
| *dolphin* | *turtle* |
| *donkey* | *whale* |
| baseball | spiders |
| bowling | sharks |
| sharks | bowling |
| spiders | golf |
| tennis | football |
| soccer | tennis |
| volleyball | baseball |
| golf | soccer |
| basketball | volleyball |
| softball | softball |
| football | basketball |

TABLE IV COMPARISON OF RANKING IN THE CONTEXT VENUS BETWEEN SIMPLE PMING DISTANCE AND PMING DISTANCE WITH VECTOR SPACE MODEL (VSM) FROM [6]

| PMING | PMING(VSM) |
|---|---|
| **venus** | **venus** |
| **saturn** | **saturn** |
| **moon** | **moon** |
| *tennis* | **space** |
| **space** | wolf |
| wolf | spiders |
| snake | snake |
| *baseball* | horses |
| *soccer* | dolphin |
| horses | donkey |
| bears | whale |
| dolphin | turtle |
| whale | bears |
| spiders | sharks |
| *basketball* | *golf* |
| sharks | *football* |
| turtle | *bowling* |
| *golf* | *tennis* |
| *volleyball* | *volleyball* |
| *football* | *soccer* |
| donkey | *softball* |
| *softball* | *baseball* |
| *bowling* | *basketball* |